\documentclass{article}

\usepackage{amsmath,amssymb}    %%一些数学符号
\usepackage{color}
\usepackage{cite}               %%是引用格式变漂亮
\usepackage{hyperref}           %%超链接必须

\numberwithin{equation}{section}   %%公式按节编号

\def \be {\begin{equation}}
\def \ee {\end{equation}}
\def \ba {\begin{array}}
\def \ea {\end{array}}
\def \bea{\begin{eqnarray}}
\def \eea{\end{eqnarray}}

\def \a {\alpha}
\def \b {\beta}
\def \g {\gamma}

\def \G {\Gamma}
\def \d {\delta}

\def \e {\epsilon}

\def \m {\mu}
\def \n {\nu}
\def \k {\kappa}
\def \l {\lambda}
\def \L {\Lambda}
\def \s {\sigma}
\def \S {\Sigma}
\def \r {\rho}
\def \o {\omega}
\def \O {\Omega}

\def \t {\tau}

\def \p {\partial}
\def \f {\frac}
\def \na {\nabla}

\def \nn {\nonumber}

\def \mc {\mathcal}
\def \mf {\mathfrak}

\def \lt {\left}
\def \rt {\right}

\def \sr {\sqrt}
\def \td {\tilde}

\begin{document}

\title{\textbf{Note on generalized gravitational entropy in Lovelock gravity}}
\author{
Bin Chen$^{1,2}$\footnote{bchen01@pku.edu.cn}\,
and
Jia-ju Zhang$^{1}$\footnote{jjzhang@pku.edu.cn}
}
\date{}

\maketitle

\begin{center}
{\it
$^{1}$Department of Physics and State Key Laboratory of Nuclear Physics and Technology, Peking University, No.~5 Yiheyuan Road, Beijing 100871, P.~R. China\\
\vspace{2mm}
$^{2}$Center for High Energy Physics, Peking University, No.~5 Yiheyuan Road, Beijing 100871, P.~R. China
}
\vspace{10mm}
\end{center}

\begin{abstract}

The recently proposed gravitational entropy generalize the usual black hole entropy to Euclidean solutions without $U(1)$ symmetry in the framework of Einstein gravity. The entropy of such smooth configuration is given by the area of minimal surface, therefore explaining the Ryu-Takayanagi formula of holographic entanglement entropy. In this note we investigate the generalized gravitational entropy for general Lovelock gravity in arbitrary dimensions. We use the replica trick and consider the Euclidean bulk spacetime with conical singularity localized at a codimension two surface. We obtain a constraint equation for the surface by requiring the bulk equation of motion to be of  good behavior. When the bulk spacetime is maximally symmetric, the constraints show that the traces of the extrinsic curvatures of the surface are vanishing, i.e. the surface has to be geometrically a minimal surface.   However the constraint equation cannot be obtained by the variation of the known functional for holographic entanglement entropy in Lovelock gravity.

\end{abstract}

\baselineskip 18pt
\thispagestyle{empty}

\newpage

\section{Introduction}

One of the most curious characteristics of black hole thermodynamics is the area law of the Bekenstein-Hawking entropy \cite{Bekenstein:1973ur,Hawking:1974sw}
\be
S=\f{\mc A}{4G},
\ee
with $\mc A$ being the area of the black hole event horizon. This entropy was reconsidered in \cite{Gibbons:1976ue} from the point of view of Euclidean quantum gravity.  The Euclideanized black hole solution is a saddle point of the action and its thermodynamic partition function allows one to read the above entropy. In this case, the solution and the boundary condition have a $U(1)$ symmetry. Very recently in a remarkable paper \cite{Lewkowycz:2013nqa} the gravitational entropy was generalized to the situation without $U(1)$ symmetry\footnote{Note that there was similar suggestion in \cite{Fursaev:2007sg}.}. The basic setup is to consider metrics ending on a boundary. The boundary is of a noncontractible circle $\tau \sim \tau +2\pi$ and the boundary data respects the periodic condition. Moreover the spacetime in the interior has to be smooth. Then one may use the replica trick to compute entropy
\be
S=-n\p_n(\log Z(n)-n\log Z(1))|_{n=1}.
\ee
Here $\log Z(1)$ is the Euclidean action of original configuration and $\log Z(n)$ is the one of the configuration with the same boundary data but new circle $\tau \sim \tau +2\pi n$. In practice, one may analytically continue the geometries away from integer values of $n$ and consider the case $n$ is very near 1. This would introduce conical singularity localized at a codimension two surface. In this case, the net result is similar to that one introduces a very light codimension three brane into the original geometry and treats the codimension two surface as the worldvolume of the codimension three brane, and gravitational entropy comes from the Nambu-Goto action of the brane. By minimizing the Nambu-Goto action, one finds that worldvolume of the brane should respect the minimal area condition.\footnote{In the four-dimensional gravity case this would be a cosmic string. Note that there is some abuse of terminology here, because the codimension two surface is  different from the worldvolume of a codimension three brane. For example the worldvolume of a codimension three brane would include the direction of time, but the codimension two surface is not extended in the time direction.} In other words, the generalized gravitational entropy is given by the area of the minimal codimension two surface in the bulk
\be\label{e2}
S=\frac{A_{\mbox{\tiny minimal}}}{4G_N}.
\ee
In the case that the configurations have $U(1)$ symmetry, this entropy is reduced to the Bekenstein-Hawking entropy.

This generalized entropy is defined in quite general situations, basing on the holographic nature of quantum gravity. In the case that the boundary is asymptotically AdS, the boundary theory is better understood. From AdS/CFT correspondence \cite{Maldacena:1997re,Gubser:1998bc,Witten:1998qj}, the gravitational entropy defined above provides a holographic way to compute the entanglement entropy of corresponding density matrix in the boundary theory. It actually explains the Ryu-Takayanagi formula of holographic entanglement entropy \cite{Ryu:2006bv,Ryu:2006ef}.\footnote{For earlier efforts to explain Ryu-Takayanagi formula, please see \cite{Fursaev:2006ih,Fursaev:2007sg,Headrick:2010zt,Casini:2011kv,Fursaev:2012mp}.}

In fact, the above discussion is in the framework of Einstein gravity. It would be interesting to generalize the discussion to the gravity theory with higher curvature terms. As the first step, in this note we investigate Lovelock gravity in various dimensions. One virtue of Lovelock gravity is that the gravitational equations of motion involve only second order derivatives so that the study is relatively easy. Moreover we have the results on holographic entanglement entropy in Lovelock gravity as guide.

Just like the black hole entropy in higher curvature gravities \cite{Jacobson:1993xs,Wald:1993nt,Jacobson:1993vj,Iyer:1994ys}, the formula of holographic entanglement entropy for higher curvature gravities should be modified \cite{Iwashita:2006zj,Fursaev:2006ih,deBoer:2011wk,Hung:2011xb,Ogawa:2011fw}. The area law should be replaced by the extreme of another functional of the embedding of a codimension two surface in the bulk, and this is crucial for the holographic proof of the strong subadditivity of entanglement entropy \cite{Headrick:2007km}.
The direct choice is the functional given by Wald formula \cite{Wald:1993nt,Jacobson:1993vj,Iyer:1994ys}. However this choice is spoiled by the ambiguities of the extrinsic curvatures of the embedding \cite{Fursaev:2006ih,deBoer:2011wk}. It was argued in \cite{Hung:2011xb} that in general the Wald formula is not the correct one. Instead it was suggested that  the functional for Lovelock gravity could be given by the Jacobson-Myers formula\cite{Jacobson:1993xs}, which is purely determined by the intrinsic curvature of the submanifold.   The Jacobson-Myers functional was first proposed as  the black hole entropy formula, and also was suggested in \cite{Iyer:1994ys} as the entropy of a dynamic black hole in Lovelock gravity\footnote{This was pointed out explicitly in \cite{Lewkowycz:2013nqa}.}. It gives the same result as the Wald formula for the black hole. The proposal of \cite{Hung:2011xb} was applied to AdS solitons in \cite{Ogawa:2011fw} and well-behaved results were obtained. There was also investigation on the thermodynamics of entanglement entropy using holographic method for Lovelock gravity in \cite{Guo:2013aca}, which generalizes the analysis in \cite{Bhattacharya:2012mi,Nozaki:2013vta} for Einstein gravity.

For the generalized gravitational entropy, it is not a priori clear what kind of functional we should use. There are a few consistent checks on the action functional. First of all it must reduce to the Wald formula for the black hole. Secondly it may give the holographic entanglement entropy when the asymptotic boundary is AdS. However the basic requirement for the functional is that its variation should give the constraint equation of the codimension two surface.
In the present work we use the method in \cite{Lewkowycz:2013nqa} to find the constraint equation for the codimension two surface in Lovelock gravity, and compare it with the one from the variation of functional law for holographic entanglement entropy in Lovelock gravity as suggested in \cite{Hung:2011xb}. We find that they disagree with each other. If the result in this note is correct, it suggests either that the generalized gravitational entropy in Lovelock gravity should not be interpreted as the holographic entanglement entropy or the Jacobson-Myers functional in \cite{Hung:2011xb} needs modification. Also it is possible that the procedure in this paper should be modified to yield desirable results, and we will discuss this further in the conclusion section of the note. For comparison we also consider the variation of the Wald functional when the bulk spacetime has maximal symmetry, and we find that the constraint  equation  cannot be got from the Wald functional either.

The remaining part of the paper is organized as follows. In Section~\ref{s2} we consider the Gauss-Bonnet gravity. We firstly get the constraint equation for the codimension two surface from the equation of motion of the bulk with conical singularity, and then we try to get the equation of the surface from the variation of the Jacobson-Myers functional. We find that two equations disagree with each other. In Section~\ref{s3} we do similar analysis for general Lovelock gravity. We end with conclusion and discussion in Section~\ref{s5}. There are some calculation details in Appendix~\ref{sa} and \ref{sb}.

{\bf Note：} At the same time the first version of this paper appeared in arXiv, there appeared another paper \cite{Bhattacharyya:2013jma} which has some overlaps with our work. The authors in \cite{Bhattacharyya:2013jma} considered holographic entanglement entropy in five-dimensional Gauss-Bonnet gravity and the results there are not in conflict with ours. However there were additional approximations used there, based on which it was claimed that the constraint equation from conical geometry agreed with that from the variation of Jacobson-Myers functional. This inspires us to investigate in detail the difference between the constraint equation from conical geometry with the one from the variation of Jacobson-Myers functional. We find that the difference is negligible if the cubic terms is much smaller than the linear terms (\ref{j45}). We present our investigations in Section~\ref{s4}

\section{Gauss-Bonnet gravity}\label{s2}

For simplicity, in this section we consider the generalized gravitational entropy in Gauss-Bonnet gravity.

\subsection{Black hole entropy}

We use Euclidean signature for the $(d+1)$-dimensional spacetime manifold $\mc M$, and the coordinates are $x^\m$ and the metric is $g_{\m\n}$. The action of Gauss-Bonnet gravity has the form
\be \label{gb}
I_{GB}=-\f{1}{16\pi G}\int_\mc M d^{d+1}x\sr{g}\lt[ R-2\L +\l (R^2-4R_{\m\n}R^{\m\n}+R_{\m\n\r\s}R^{\m\n\r\s}) \rt],
\ee
where the boundary terms and possible matter actions have been omitted. The equation of motion is
\bea \label{eomgb}
&& R_{\m\n}+2\l (RR_{\m\n}-2R_{\m\r}R_\n^{\phantom{\n}\r}-2R^{\r\s}R_{\r\m\s\n}+R_{\m\r\s\l}R_\n^{\phantom{\n}\r\s\l}) \nn\\
&& -\f{1}{2} g_{\m\n} \lt[ R-2\L+\l(R^2-4R_{\r\s}R^{\r\s}+R_{\r\s\l\t}R^{\r\s\l\t}) \rt]=8\pi G T_{\m\n}
\eea
with $T_{\m\n}$ being the energy-momentum-stress tensor of matter.

To get the entropies of black holes in a higher curvature gravity, one can use the Wald formula \cite{Wald:1993nt,Jacobson:1993vj,Iyer:1994ys}, or equivalently the conical singularity method \cite{Solodukhin:1994yz,Fursaev:1995ef}.  From the Euclidean gravity action
\be
I=-\int_\mc M d^{d+1}x\sr{g} L,
\ee
the Wald formula of the black hole entropy is
\be
S=2\pi \int_\S d^{d-1}y \sr{h} \f{\p L}{\p R_{\m\n\r\s}} \e_{\m\n}\e_{\r\s},
\ee
with $\S$ being the event horizon, $y^i$ and $h_{ij}$ being the coordinates and metric of $\S$. The horizon is of codimension two.  There are two normal vectors $n_{(\a)}^\m$ with $\a=1,2$, being normalized as $n^\m_{(\a)}n_{(\b)\m}=\d_{\a\b}$, from which one may define
\be
\e^{\m\n}=n_{(1)}^\m n_{(2)}^\n-n_{(2)}^\m n_{(1)}^\n.
\ee
It can be checked easily that \cite{Azeyanagi:2007bj,Myers:2010tj}
\be
\e^{\m\n}\e^{\r\s}=(n^\m n^\r)(n^\n n^\s)-(n^\m n^\s)(n^\n n^\r),
\ee
with the definition $(n^\m n^\n)=\sum_{\a=1}^2 n_{(\a)}^\m n_{(\a)}^\n$. There are useful relations $(n^\m n^\n)(n_\n n^\r)=(n^\m n^\r)$ and $(n^\m n_\m)=2$.

Using the Wald formula, for Gauss-Bonnet gravity one can get the entropy
\be \label{f2}
S_{GB}=\f{1}{4G}\int_\S d^{d-1}y \sr{h} \lt[ 1+2\l(R-2R_{\m\n}(n^\m n^\n)+R_{\m\n\r\s}(n^\m n^\r)(n^\n n^\s)) \rt].
\ee
From Gauss-Codazzi equation one has
\be
\mc R=R-2R_{\m\n}(n^\m n^\n)+R_{\m\n\r\s}(n^\m n^\r)(n^\n n^\s)+K_{(\a)}K_{(\a)}-K_{(\a)\m\n}K_{(\a)}^{\m\n},
\ee
where $\mc R$ is the intrinsic curvature of $\S$ and summation of the index $\a$ is indicated. The extrinsic curvatures and their traces are defined as
\bea \label{e34}
&& K_{(\a)\m\n}=h_{(\m}^\r h_{\n)}^\s \na_\r n_{(\a) \s}  ,\nn\\
&& K_{(\a)}=h^{\m\n}K_{(\a)\m\n},
\eea
with the induced metric being
\be
h_{\m\n}=g_{\m\n}-(n_\m n_\n).
\ee
One can also define the projected Riemann tensor, Ricci tensor and Ricci scalar as
\bea
&& \mf R_{\m\n\r\s} \equiv h_\m^\l h_\n^\t h_\r^\k h_\s^\o R_{\l\t\k\o},  \nn\\
&& \mf R_{\m\n} \equiv h^{\r\s} \mf R_{\r\m\s\n}, \nn\\
&& \mf R \equiv h^{\m\n} \mf R_{\m\n}.
\eea
Then the entropy of the black hole in Gauss-Bonnet gravity is
\be \label{e24}
S_{GB}=\f{1}{4G}\int_\S d^{d-1}y \sr{h} \lt( 1+2\l \mf R \rt).
\ee
Since the extrinsic curvatures vanish for the black hole horizon, this is just
\be
S_{GB}=\f{1}{4G}\int_\S d^{d-1}y \sr{h} \lt( 1+2\l \mc R \rt).
\ee

\subsection{Constraint equation from replica trick}

In this subsection we use the arguments in \cite{Lewkowycz:2013nqa} with some modifications to discuss the generalized gravitational entropy in the Gauss-Bonnet gravity. We see how the equation of motion for Gauss-Bonnet gravity in the bulk with conical singularity localized at a codimension two surface constrain the surface. Note that as stated in \cite{Lewkowycz:2013nqa} for Einstein gravity, the constraint equation is similar to the equation of motion for the world volume of a codimension three brane; i.e. a $(d-2)$-brane in $(d+1)$-dimensional spacetime $\mc M$. When $\mc M$ is four-dimensional it would be similar to the equation for the world sheet of a cosmic string \cite{Unruh:1989hy,Boisseau:1996bp}.

For a $(d+1)$-dimensional manifold $M$, at the vicinity of a codimension two surface $\S$, we could have the approximation of metric  locally
\be \label{e8}
ds^2=g_{\a\b}dx^\a dx^\b+(h_{ij}+2x^\a K_{(\a)ij})dy^idy^j+\cdots,
\ee
with $\a=1,2$, and $i=1,2,\cdots,d-1$. Here we also have
\be
g_{\a\b}dx^\a dx^\b=(dx^1)^2+(dx^2)^2=dr^2+r^2d\phi^2=dz d\bar z.
\ee
The notation $\cdots$ represents the deviations from our approximation, which is supposed not to contribute in our analysis.
In (\ref{e8}) $K_{(\a)ij}$ is just the extrinsic curvature for the embedding of $\S$ in $\mc M$, and $h_{ij}$ is  the metric of $\S$. In the replica trick, there is a conical singularity localized everywhere on $\S$ \cite{Fursaev:2006ih}. We makes $n$ copies of the bulk and identify them properly, and when $n=1+\e$ with $\e$ being infinitesimally small the metric (\ref{e8}) becomes
\be \label{cone}
ds^2=e^{2\r}g_{\a\b}dx^\a dx^\b+(h_{ij}+2x^\a K_{(\a)ij})dy^idy^j+\cdots,
\ee
with
\be \r=-\e\ln r=-\f{\e}{2}\ln(z\bar z). \ee
Note that the terms we omitted in (\ref{e8}) would be modified too, and so the $\cdots$ in (\ref{e8}) and (\ref{cone}) would be different, but we expect that such modification will not affect our final result\footnote{We thank Dmitri Fursaev for discussion on this issue.}.

Now we would like to investigate the equation of motion for the Gauss-Bonnet gravity (\ref{eomgb}) for the conical metric (\ref{cone}). We  only consider the vicinity of the surface, i.e. near the origin of the cone, to  the leading order of $\e$. We focus on the divergent terms of order $\p_\a \r \sim \f{1}{r}$, and do not care the $\d$-function terms $\p^a\p_\a\r \sim \d^2(x)$ or the terms $\r\sim\ln r$. Note that in the following part of this section what we mean by `equals' is the equality with the above approximations in mind, the subleading terms and the terms that do not contribute to order $\f{1}{r}$ in the final results are omitted.

Firstly we have to calculate the curvature tensors for the metric (\ref{cone}). The results are summarized in Appendix~\ref{sa}. With all the results, from the ($zz$)-component of the equation of motion (\ref{eomgb}), we get
\be
8\pi T_{zz}=2\p_z \r \lt[ h^{ij}-4\l \lt( \mf R^{ij}-\f{1}{2}\mf R h^{ij} \rt) \rt] K_{(z)ij}+\cdots,
\ee
with $\cdots$ being terms that do not contribute at order $\f{1}{r}$.
We suppose $T_{zz}$ is well-behaved near $r=0$, so we have
\be
(h^{ij}-4\l \mf G^{ij})K_{(z)ij}=0,
\ee
with $\mf G^{ij}=\mf R^{ij}-\f{1}{2}\mf R h^{ij}$. Similarly from the ($\bar z \bar z$)-component of the equation of motion we get
\be
(h^{ij}-4\l \mf G^{ij})K_{(\bar z)ij}=0.
\ee
In summary, we have
\be \label{gbrt}
(h^{ij}-4\l \mf G^{ij})K_{(\a)ij}=0, ~~ \a=1,2.
\ee

When the bulk spacetime $\mc M$ has maximal symmetries
\be \label{e30}
R_{\m\n\r\s}=\f{2\td \L}{d(d-1)}(g_{\m\r}g_{\n\s}-g_{\m\s}g_{\n\r}),
\ee
we have
\be
\mf R_{ijkl}=\f{2 \td \L}{d(d-1)}( h_{ik} h_{jl}-h_{il} h_{jk}).
\ee
Then the constraint equation (\ref{gbrt}) becomes
\be \label{e27}
\lt( 1+\f{4(d-2)(d-3)}{d(d-1)} \l \td\L \rt) K_{(\a)}=0, ~~ \a=1,2.
\ee
which is just the condition that the trace of extrinsic curvature is vanishing
\be
K_{(\a)}=0, ~ \a=1,2.
\ee
In other words, for maximally symmetric spactime, the codimension two surface must be
a minimal surface. % however we write it in this clumsy way for later comparison.

\subsection{Functional law?} \label{s22}

In \cite{Hung:2011xb}, it was argued that for the Gauss-Bonnet gravity, as well as general Lovelock gravity which we will consider in the next section, the functional for the holographic entanglement entropy should be given solely by the intrinsic curvature of the surface and some boundary terms for the variation problem being well defined. Explicitly, for Gauss-Bonnet gravity the functional is unambiguously given by
\be \label{e3}
S_{GB}=\f{1}{4G}\int_\S d^{d-1}y \sr{h}(1+2\l \mc R)+\f{\l}{G}\int_{\p\S} d^{d-2}y \sr{\s}\mc K,
\ee
with $\mc K$ being the trace of the extrinsic curvature for the embedding of $\p\S$ in $\S$. As is known, the extreme of the area law is equivalent to the vanishing of the trace of the extrinsic curvatures, and one could see, for example, in \cite{Hubeny:2007xt} for a derivation. In this section we will show what the functional (\ref{e3}) leads to.

 We start from a $D$-dimensional Euclidean manifold $\mc M$ with coordinates $x^\m$ and metric $g_{\m\n}$. We consider a codimension $n$ surface $\S$ embedded in $\mc M$, and we label its coordinates by $y^i$. For the surface $\S$ there would be $n$ normal vectors $n_{(\a)}^\m$ with $\a=1,2,\cdots,n$. The normal vectors are chosen such that
\be
n_{(\a)}^\m n_{(\b)\m}=\d_{\a\b}.
\ee
The induced metric on $\S$ is defined as
\be
h_{\m\n}=g_{\m\n}-\sum_{\a=1}^n n_{(\a)\m} n_{(\a)\n}.
\ee
and the extrinsic curvatures and their traces are defined the same as (\ref{e34}).

The embedding of $\S$ in $\mc M$ could be characterized by the functions
\be \label{e5}
x^\m=X^\m(y).
\ee
The vector $\p_i X^\m$ is the tangent to the $\S$, and
\be
\p_i X^\m n_{(\a)\m}=0,
\ee
for arbitrary $i=1,2,\cdots,D-n$ and $\a=1,2,\cdots,n$. The metric $h_{ij}$ on $\S$ is the pullback of $g_{\m\n}$ on $\mc M$,
\be
h_{ij}=\p_i X^\m \p_j X^\n g_{\m\n}=\p_i X^\m \p_j X^\n h_{\m\n}.
\ee
There is a useful relation
\be
h^{ij}\p_i X^\m \p_j X^\n=h^{\m\n}.
\ee
Other tensors on $\mc M$ could be pulled back to $\S$ too, and for example there is
\be
K_{(\a)ij} \equiv \p_i X^\m \p_j X^\n K_{(\a)\m\n}=\p_{(i} X^\m \p_{j)} X^\n \na_\m n_{(\a)\n}.
\ee

The problem is to find the equation that follows from the variation of the functional
\be \label{e6}
\mc A_{GB}=\int_\S d^{D-n}y \sr{h}(1+2\l \mc R)+4\l\int_{\p\S} d^{D-n-1}y \sr{\s}\mc K
\ee
with $\p \S$ being fixed. Note that we are varying the different embedding, and so this is just the variation of $X^\m(y)$ in (\ref{e5}). We have
\be \label{e7}
\d h_{ij}=\p_i \d X^\m \p_j X^\n g_{\m\n}
          +\p_i X^\m \p_j \d X^\n g_{\m\n}
          + \p_i X^\m\p_j X^\n \p_\r g_{\m\n}\d X^\r.
\ee
Since $\p \S$ is fixed, at the boundary we have
\be
\d X^\m|_{\p \S}=0,
\ee
and from which there is
\be
\d h_{ij}|_{\p \S}=0.
\ee
Thus the variation of (\ref{e6}) is well defined, and we get
\be
\d \mc A_{GB}=\f{1}{2}\int_\S d^{D-n}y \sr{h}(h^{ij}-4\l \mc G^{ij})\d h_{ij},
\ee
with $\mc G_{ij}$ being the Einstein tensor defined by $h_{ij}$. Note that on $\S$, $X^\m$, $g_{\m\n}$ and $n_{(\a)}^\m$ are all scalars. Then using (\ref{e7}), from $\f{\d \mc A_{GB}}{\d X^\m}=0$ we have \cite{Fursaev:2006ih}
\be
(h^{ij}-4\l \mc G^{ij})\Pi^\m_{ij}=0,
\ee
where
\be
\Pi^\m_{ij}=D_i\p_j X^\m+\G^\m_{\r\s}\p_i X^\r \p_j X^\s,
\ee
with $D_i$ being the covariant derivative with respect to $h_{ij}$ and $\G^\m_{\r\s}$ being the Christoffel connection defined by the metric $g_{\m\n}$. In deriving the above formula we have used the fact that $D_i h^{ij}=D_i \mc G^{ij}=0$. One can show that
\be
\Pi^\m_{ij} g_{\m\n} \p_k X^\n=0,
\ee
which stems from the diffeomorphism invariance of the functional (\ref{e6}). Thus the only independent  components of $\Pi^\m_{ij}$ are
\be
n_{(\a)\m}\Pi^\m_{ij}=-K_{(\a)ij},
\ee
which is just the pullback of the extrinsic curvature. Thus we could conclude that the extreme condition for the functional (\ref{e6}) is equivalent to the requirement
\be \label{gbfl}
(h^{ij}-4\l \mc G^{ij})K_{(\a)ij}=0, ~~ \a=1,2,\cdots,n.
\ee

The result here is ready to be compared to the result (\ref{gbrt}) from the replica trick. Immediately we find that they are different. In the result (\ref{gbrt}) there is the pullback curvature, and in the result (\ref{gbfl}), which follows from the proposal of \cite{Hung:2011xb}, there is the intrinsic curvature. Then we conclude that the generalized gravitational entropy for Gauss-Bonnet gravity is not the holographic entanglement entropy.

Still we are curious whether the equation (\ref{gbrt}) could be got by variation of some other functionals. Another natural candidate  would be the Wald functional (\ref{e25})
\be \label{e25}
\mf A_{GB}=\int_\S d^{D-n}y \sr{h}(1+2\l \mf R)+ \int_{\p\S} d^{D-n-1}y \sr{\s} \cdots
\ee
 Note that there are possible boundary terms that renders the variation problem being well defined, and we will not try to pursue these terms in this paper. In varying the functional (\ref{e25}) we consider only the case when the bulk space $\mc M$ has maximal symmetries. In this case we have
\be
\mf R_{\m\n\r\s}=\f{2\td \L}{(D-1)(D-2)}(h_{\m\r}h_{\n\s}-h_{\m\s}h_{\n\r}),
\ee
and then the functional (\ref{e25}) becomes
\be
\mf A_{GB}= \lt( 1+\f{4(D-n)(D-n-1)}{(D-1)(D-2)}\l\td\L  \rt)\int_\S d^{D-n}y \sr{h}+ \int_{\p\S} d^{D-n-1}y \sr{\s} \cdots.
\ee
Then the variation of the above functional is the same as the one from the area functional and leads to the result that  the traces of the extrinsic curvatures are vanishing
\be
K_{(\a)}=0, ~~ \a=1,2,\cdots,n.
\ee
However, we would like to write it in a clumsy way
\be
\lt( 1+\f{4(D-n)(D-n-1)}{(D-1)(D-2)}\l\td\L  \rt)K_{(\a)}=0, ~~ \a=1,2,\cdots,n.
\ee
For $D=d+1, n=2$, it becomes
\be \label{e26}
\lt( 1+\f{4(d-2)}{d}\l\td\L  \rt)K_{(\a)}=0, ~~ \a=1,2.
\ee

Indeed when the bulk spacetime $\mc M$ has maximal symmetries the variation of the Wald functional leads to the conclusion that the traces of the extrinsic curvatures of the embedded surface must be vanishing, the same as the constraint equation. However the prefactors in (\ref{e26}) and (\ref{e27}) are disturbingly different. Actually they are only in match when $\td \L=0$ or $d=2$. This mismatch strongly suggests that  the equation (\ref{gbrt}) could not be obtained by the variation of the functional (\ref{e25}).

\section{Lovelock gravity}\label{s3}

In this section we consider the general Lovelock gravity. The calculation is parallel to that of Gauss-Bonnet gravity, and the result and conclusion are similar.

\subsection{Black hole entropy}

The action of general Lovelock gravity could be written as
\be
I_L=-\f{1}{16\pi G} \int_{\mc M} d^{d+1}x \sr{g}\sum_{m=0}^{[\f{d+1}{2}]} \l_m L_{(m)},
\ee
with $[\f{d+1}{2}]$ being the integer part of $\f{d+1}{2}$, and $L_{(m)}$ being the $m$-th order Lagrangian of Lovelock gravity
\be \label{lm}
L_{(m)}=\f{(2m)!}{2^m}\d^{\m_1\n_1\cdots\m_m\n_m}_{\r_1\s_1\cdots\r_m\s_m}R^{\r_1\s_1}_{\m_1\n_1} \cdots R^{\r_m\s_m}_{\m_m\n_m},
\ee
where we have used the definition that
\bea
&& \d^{\m_1\n_1\cdots\m_m\n_m}_{\r_1\s_1\cdots\r_m\s_m} \equiv \d^{\m_1}_{\lt[\r_1\rt.} \d^{\n_1}_{\s_1} \cdots
                                                            \d^{\m_m}_{\r_m} \d^{\n_m}_{\lt.\s_m\rt]},  \nn\\
&& R^{\r\s}_{\m\n} \equiv R_{\m\n}^{\phantom{\m\n}\r\s}.
\eea
It can be shown that
\bea
&& L_{(0)}=1, ~~~ L_{(1)}=R, \nn\\
&& L_{(2)}=R^2-4R_{\m\n}R^{\m\n}+R_{\m\n\r\s}R^{\m\n\r\s}.
\eea
 If only keeping $\l_0=-2\L, \l_1=1$ nonvanishing, we get the Einstein gravity, while keeping $\l_0=-2\L, \l_1=1, \l_2=\l$ nonvanishing we obtain the Gauss-Bonnet gravity (\ref{gb}). The equation of motion for the Lovelock gravity is
\be \label{eoml}
\sum_{m=0}^{[\f{d+1}{2}]} \l_m \lt( P_{(m)(\m}^{\phantom{(m)(\m}\r\s\l} R_{\n)\r\s\l} -\f{1}{2}L_{(m)}g_{\m\n}  \rt)=8\pi G T_{\m\n},
\ee
where we have defined
\be
P_{(m)\r}^{\phantom{(m)\r}\s\m\n} \equiv \f{\p L_{(m)}}{\p R^\r_{\phantom{\r}\s\m\n}}.
\ee
Explicitly we have
\be
P_{(m)\r\s}^{\phantom{(m)}\m\n} \equiv P_{(m)\r\s}^{\phantom{(m)\r\s}\m\n}
  =\f{m(2m)!}{2^m}
   \d^{\m\n\m_2\n_2\cdots\m_m\n_m}_{\r\s\r_2\s_2\cdots\r_m\s_m}R^{\r_2\s_2}_{\m_2\n_2} \cdots R^{\r_m\s_m}_{\m_m\n_m}.
\ee

Using the Wald formula, the black hole entropy in the Lovelock gravity could be calculated as
\be \label{e16}
S^{(W)}_L=\f{1}{4G} \int_\S d^{d-1}y \sr{h} \sum_{m=1}^{[\f{d+1}{2}]}m\l_m \mf L_{(m-1)}.
\ee
Actually the black hole entropy could also be derived using the Hamiltonian approach \cite{Jacobson:1993xs}
\be \label{e17}
S^{(JM)}_L=\f{1}{4G} \int_\S d^{d-1}y \sr{h} \sum_{m=1}^{[\f{d+1}{2}]}m\l_m \mc L_{(m-1)}.
\ee
The quantities $\mf L_{(m-1)}$ and $\mc L_{(m-1)}$ are $(m-1)$-th order Lovelock Lagrangians defined on the horizon $\S$, and the difference is that $\mf L_{(m-1)}$ is defined from the projected curvature $\mf R_{ijkl}$ (\ref{pc}) and $\mc L_{(m-1)}$ is defined from the intrinsic curvature $\mc R_{ijkl}$ of $\S$. For the black hole horizon the extrinsic curvature vanishes so that the above two formulas coincide.

\subsection{Constraint from replica trick}

In this subsection we evaluate the equation of motion  (\ref{eoml}) for the geometry with conical singularity (\ref{cone}). Note that we only focus on the $\f{1}{r}$ terms. Firstly we consider the ($zz$)-component
\be \label{e39}
8\pi G T_{zz}=\sum_{m=0}^{[\f{d+1}{2}]} \l_m P_{(m)z}^{\phantom{(m)z}\r\s\l} R_{z\r\s\l}.
\ee
We leave the details of the calculation in Appendix~\ref{sb}. The result is
\be
8\pi G T_{zz}=-4\p_z \r K_{(z)ij} \sum_{m=1}^{[\f{d+1}{2}]} m \l_m
\lt( \mf P_{(m-1)\phantom{i}klp}^{\phantom{(m-1)}i} \mf R^{jklp} -\f{1}{2}\mf L_{(m-1)} h^{ij} \rt)+\cdots,
\ee
with the definition (\ref{e37}).

The well-behavior of $T_{zz}$ near the origin leads to
\be
\sum_{m=1}^{[\f{d+1}{2}]} m \l_m
\lt( \mf P_{(m-1)\phantom{i}klp}^{\phantom{(m-1)}i} \mf R^{jklp} -\f{1}{2}\mf L_{(m-1)} h^{ij} \rt) K_{(z)ij}=0.
\ee
This result plus similar one from the ($\bar z \bar z$)-component give the final result
\be \label{j28}
\sum_{m=1}^{[\f{d+1}{2}]}
\lt[ m\l_m \lt( \mf P_{(m-1)\phantom{i}klp}^{\phantom{(m-1)}i} \mf R^{jklp} -\f{1}{2}\mf L_{(m-1)} h^{ij} \rt) \rt]
K_{(\a)ij}=0, ~~ \a=1,2.
\ee
In the case of the bulk has maximal symmetries (\ref{e30}), the above equation becomes
\be \label{j29}
\sum_{m=1}^{[\f{d+1}{2}]} m\l_m \f{(d-2)!}{(d-2m)!}
           \lt[ \f{2\td\L}{d(d-1)} \rt]^{m-1} K_{(\a)}=0, ~~ \a=1,2.
\ee
In this case, the traces of the extrinsic curvatures of the embedded surface are vanishing.

\subsection{Functional}

In \cite{Hung:2011xb}, it was argued that the functional is given by (\ref{e17}) plus some boundary terms for the variational problems to be well defined.  The boundary terms are not guaranteed to exist for general higher curvature gravity, but they do exist for Lovelock gravity \cite{Bunch:1981,Myers:1987yn}. From the variation of the Jacobson-Myers functional
\be
\mc A_{L}=\int_\S d^{D-n}y \sr{h}\sum_{m=1}^{[\f{D-n}{2}]+1}m\l_m \mc L_{(m-1)}+\int_{\p\S} d^{D-n-1}y \sr{\s}\cdots,
\ee
we get
\be
\d \mc A_{L} = \f{1}{2} \int_\S d^{D-n} y \sr{h} \sum_{m=1}^{[\f{D-n}{2}]+1} \lt[ m \l_m
               \lt( \mc L_{(m-1)} h^{ij}-2 \mc P_{(m-1)\phantom{i}klp}^{\phantom{(m-1)}i} \mc R^{jklp} \rt)\rt] \d h_{ij},
\ee
with
\be
\mc P_{(m-1)i}^{\phantom{(m-1)i}klp} \equiv \f{\p \mc L_{(m-1)}}{\p \mc R^i_{\phantom{i}klp}}.
\ee
Then using
\be
D_i \lt( \mc P_{(m-1)\phantom{(i}klp}^{\phantom{(m-1)}(i} \mc R^{j)klp} -\f{1}{2}\mc L_{(m-1)} h^{ij} \rt)=0,
\ee
we arrive at
\be
\sum_{m=1}^{[\f{D-n}{2}]+1}
\lt[ m\l_m \lt( \mc P_{(m-1)\phantom{i}klp}^{\phantom{(m-1)}i} \mc R^{jklp} -\f{1}{2}\mc L_{(m-1)} h^{ij} \rt) \rt]
\Pi^\m_{ij}=0,
\ee
which is equivalent to
\be \label{fll}
\sum_{m=1}^{[\f{D-n}{2}]+1}
\lt[ m\l_m \lt( \mc P_{(m-1)\phantom{i}klp}^{\phantom{(m-1)}i} \mc R^{jklp} -\f{1}{2}\mc L_{(m-1)} h^{ij} \rt) \rt]
K_{(\a)ij}=0, ~~ \a=1,2,\cdots,n.
\ee

Similarly, we may consider the variation of Wald  functional
\be \label{e33}
\mf A_{L}=\int_\S d^{D-n}y \sr{h}\sum_{m=1}^{[\f{D-n}{2}]+1}m\l_m \mf L_{(m-1)}+\int_{\p\S} d^{D-n-1}y \sr{\s}\cdots.
\ee
Again we only consider the simple case in which the bulk spacetime is of maximal symmetries, and then the above functional is
\be
\mf A_{L}= \sum_{m=1}^{[\f{D-n}{2}]+1}m\l_m \f{(D-n)!}{(D-n-2m+2)!}
           \lt[ \f{2\td\L}{(D-1)(D-2)} \rt]^{m-1}
           \int_\S d^{D-n}y \sr{h}+\int_{\p\S} d^{D-n-1}y \sr{\s}\cdots.
\ee
Then the variation of the functional for the special case $D=d+1, ~ n=2$ leads to
\be \label{j32}
\sum_{m=1}^{[\f{d+1}{2}]} m\l_m \f{(d-1)!}{(d+1-2m)!}
           \lt[ \f{2\td\L}{d(d-1)} \rt]^{m-1} K_{(\a)}=0, ~~ \a=1,2.
\ee

Obviously the relation (\ref{j28}) is different from (\ref{fll}). This means that the constraint equation for the codimension two surface (\ref{j28}) could not be found by the variation of the functional (\ref{e17}). Also the prefactor of (\ref{j29}) is different from that of (\ref{j32}), and this means that the equation (\ref{j28}) could not be got by the variation of the functional (\ref{e33}) either.

\section{Further investigation}\label{s4}

Due to the intimate relation between generalized gravitation entropy and holographic entanglement entropy, even if the Jacobson-Myers functional may not be the functional for the gravitational entropy, it may be the dominant part. It would be interesting to compare the constraint equation from generalized gravitational gravity and the equation got from Jacobson-Myers functional in the case that the explicit computation is available. In this section we make such comparison for the cases that the entangling surface is a sphere and a cylinder and the bulk is pure AdS$_5$ in Gauss-Bonnet gravity.

As is shown before for Gauss-Bonnet gravity the constraint equation of the codimension two surface from the bulk conical geometry is (\ref{gbrt})
\be \label{j39}
\lt[ h^{ij}-4\l \lt( \mf R^{ij}-\f{1}{2}\mf R h^{ij} \rt) \rt] K_{(\a)ij}=0, ~~ \a=1,2,
\ee
and that equation from the variation of the Jacobson-Myers functional is (\ref{gbfl})
\be \label{j40}
\lt[ h^{ij}-4\l \lt( \mc R^{ij}-\f{1}{2}\mc R h^{ij} \rt) \rt] K_{(\a)ij}=0, ~~ \a=1,2.
\ee
Note that $\mf R_{ij}, \mf R$ are the projected curvatures, and $\mc R_{ij}, \mc R$ are the intrinsic curvatures of $\S$.
From Gauss-Codazzi equation there are (\ref{j43})
\bea
&& \mf R_{ijkl}=\mc R_{ijkl}-K^{(\a)}_{ik} K_{(\a)jl}+K^{(\a)}_{il} K_{(\a)jk},  \nn\\
&& \mf R_{ij}=\mc R_{ij}-K^{(\a)}K_{(\a)ij}+K^{(\a)}_{ik}K_{(\a)j}^{\phantom{(\a)}k},  \nn\\
&& \mf R=\mc R-K^{(\a)}K_{(\a)}+K^{(\a)}_{ij}K_{(\a)}^{ij}.
\eea
Then the difference between the two equations (\ref{j39}) and (\ref{j40}) is
\be \label{j41}
4 \l
\lt[ ( K^{(\b)} K_{(\b)ij} -K^{(\b)}_{ik} K_{(\b)j}^{\phantom{(\a)}k} )-\f{1}{2}h_{ij} (K^{(\b)}K_{(\b)}-K^{(\b)kl}K_{(\b)kl}) \rt]
K_{(\a)}^{ij},
~~ \a=1,2.
\ee
In the case of static geometry, we can choose the $x^1$ coordinate in the conical geometry (\ref{cone}) to be the Euclidean time $x^1=t$, thus we have
\be
K_{(1)ij}=0.
\ee
Then the difference (\ref{j41}) becomes
\be \label{j42}
-2\l \lt[ K_{(2)} K_{(2)} K_{(2)}  -3K_{(2)} K_{(2)ij} K_{(2)}^{ij}  +2K_{(2)ij} K_{(2)}^{jk} K_{(2)k}^{\phantom{(2)}i} \rt],
\ee
which is in accord with the result in \cite{Bhattacharyya:2013jma}. If these cubic terms of the extrinsic curvatures are much smaller than the linear terms, i.e. that
\be \label{j47}
\l \lt[ K_{(2)} K_{(2)} K_{(2)}  -3K_{(2)} K_{(2)ij} K_{(2)}^{ij}  +2K_{(2)ij} K_{(2)}^{jk} K_{(2)k}^{\phantom{(2)}i} \rt]
\ll
\{K_{(2)}, \mc G^{ij} K_{(2)ij} \},
\ee
or a little stronger condition
\be \label{j45}
\l \{ K_{(2)} K_{(2)} K_{(2)},    K_{(2)} K_{(2)ij} K_{(2)}^{ij},    K_{(2)ij} K_{(2)}^{jk} K_{(2)k}^{\phantom{(2)}i}  \}
\ll
\{K_{(2)}, \mc G^{ij} K_{(2)ij} \},
\ee
the difference (\ref{j42}) could be neglected.

Note that there are some differences between our approximations with those in \cite{Bhattacharyya:2013jma}. The approximations there are for the conical geometry, but our approximations are for the original regular geometry. For the consistency of the constraint equations (\ref{j39}) and (\ref{j40}) we do not need every linear term of the extrinsic curvature be much smaller than the cubic terms, and for Gauss-Bonnet gravity in static spacetime the requirements (\ref{j47}) or (\ref{j45}) would be enough.

When the bulk is pure AdS$_5$ in Gauss-Bonnet gravity and for spherical entangling surface in the boundary, we write the bulk geometry as
\be
ds^2=\f{\td L^2}{z^2}(dt^2+dz^2+dr^2+r^2 d\O_2^2).
\ee
The sphere in the boundary has radius $R$, and the minimal area codimension two surface $\S$ in the bulk is \cite{Ryu:2006ef}
\be
t=t_0, ~~~ z^2+r^2=R^2,
\ee
with $t_0$ being a constant. The surface has two normal vectors
\bea
&& n_{(1)\m}=\f{\td L}{z}(1,0,0,0,0), \nn\\
&& n_{(2)\m}=-\f{\td L}{z \sr{z^2+r^2}}(0,z,r,0,0),
\eea
and then the two extrinsic curvarues are vanishing
\be
K_{(1)ij}=K_{(2)ij}=0.
\ee
Thus the surface would also be the solution to both (\ref{j39}) and (\ref{j40}).

When the entangling surface is a cylinder with radius $R$, we write the bulk pure AdS$_5$ geometry as
\be
ds^2=\f{\td L^2}{z^2}(dt^2+dz^2+dx^3+dr^2+r^2 d\phi^2).
\ee
Now the minimal area surface $\S$ is
\be
t=t_0, ~~~ r=f(z),
\ee
and $f(z)$ could be got perturbatively \cite{Solodukhin:2008dh,Hung:2011xb}
\be \label{j46}
f(z)=R-\f{1}{4R}z^2+\mc O(z^4).
\ee
The metric on $\S$ is
\be
ds^2=\f{\td L^2}{z^2} \lt[ (1+f'(z)^2)dz^2+dx^2+f(z)^2 d\phi^2 \rt].
\ee
The two normal vectors of $\S$ are
\bea
&& n_{(1)\m}=\f{\td L}{z}(1,0,0,0,0), \nn\\
&& n_{(2)\m}=\f{\td L}{z \sr{1+f'(z)^2}}(0,f'(z),0,-1,0),
\eea
and the extrinsic curvatures are
\bea \label{j44}
&& K_{(1)ij}=0,  \nn\\
&& K_{(2)zz}=\mc O(z), ~~ K_{(2)zx}=\mc O(z^2), ~~ K_{(2)z\phi}=\mc O(z^2), \nn\\
&& K_{(2)xx}=\f{\td L}{2Rz}+\mc O(z),  ~~ K_{(2)x\phi}=0, \nn\\
&& K_{(2)\phi\phi}=-\f{\td L R}{2z}+\mc O(z).
\eea
Then there is
\be
K_{(2)}=\mc O(z^3).
\ee
The Einstein tensor on $\S$ is
\bea
&& \mc G_{zz}=\f{1}{z^2}+\f{1}{2R^2}+\mc O(z^2),  \nn\\
&& \mc G_{zx}=\mc O(z^3), ~~~ \mc G_{z\phi}=\mc O(z^3),  \nn\\
&& \mc G_{xx}=\f{1}{z^2}+\mc O(z^2), ~~~ \mc G_{x\phi}=\mc O(z^4),  \nn\\
&& \mc G_{\phi\phi}=\f{R^2}{z^2}-\f{1}{2}+\mc O(z^2),
\eea
from which we get
\be
\mc G^{ij}K_{(2)ij}=\mc O(z^3).
\ee
There are also
\bea
&& K_{(2)}K_{(2)}K_{(2)}=\mc O(z^9),  \nn\\
&& K_{(2)} K_{(2)ij} K_{(2)}^{ij}=\mc O(z^5),  \nn\\
&& K_{(2)ij} K_{(2)}^{jk} K_{(2)k}^{\phantom{(2)}i}=\mc O(z^5).
\eea
Thus (\ref{j46}) is solution to both (\ref{j39}) and (\ref{j40}). For large $z$ there is still possible difference, but the difference will not contribute to the divergent terms of the entanglement entropy.

For the sphere and cylinder cases in pure AdS$_5$ of Gauss-Bonnet gravity, the constraint equations from replica trick and Jacobson-Myers functional are consistent, and the results do not contradict those in \cite{Hung:2011xb}. It is tempting to conjecture that the same thing happens in general Lovelock gravity in arbitrary dimensions. Indeed the difference of the constraint equations (\ref{j28}) and (\ref{fll}) are proportional to the cubic and higher order of the extrinsic curvatures. But we can not draw such conclusion without explicit checks as above. Generally we expect that the constraint equations from replica trick and Jacobson-Myers functional for general Lovelock gravity in arbitrary dimensions would be different.

\section{Conclusion and discussion}\label{s5}

In this paper we followed the work \cite{Lewkowycz:2013nqa} and investigated the generalized gravitational entropy in Lovelock gravity. We considered general Lovelock gravity in $(d+1)$-dimensional Euclidean spacetime $\mc M$. In using the replica method, there is a conical singularity in $\mc M$ localized on a codimension two hyperspace. We required that the energy-momentum tensor to be well behaved near the cone, and obtained a constraint equation for $\S$. When the bulk spacetime is maximally symmetric, the constraint requires the vanishing of the traces of the extrinsic curvatures of the surface, or equivalently the surface should be geometrically a minimal surface.

As $\S$ is the surface where the gravitational entropy was calculated,  the constraint equation should follow from the variation of some functional of the embedding of $\S$ in $\mc M$. For Einstein gravity, this functional is just the Nambu-Goto action of the surface. For Lovelock gravity, this functional should be modified. There are two candidates for the functional. One is the Jacobson-Myers functional suggested in \cite{Hung:2011xb} which depends purely on the intrinsic curvature of the surface, and the other is the Wald functional. In \cite{Hung:2011xb} it was argued that the Wald functional is incorrect. We varied the Jacobson-Myers functional, and found that the resulting equation was not the one we got from the bulk equation of motion. We also varied the Wald functional  when the bulk spacetime has maximal symmetries, and found disagreement as well.

If one believes that the functional in \cite{Hung:2011xb} is the correct one for holographic entanglement entropy in Lovelock gravity, then the generalized gravitational entropy suggested in \cite{Lewkowycz:2013nqa} could not be taken as the holographic entanglement entropy. On the other hand if one take the generalized gravitational entropy as the holographic entanglement entropy, then one should find what was omitted in \cite{Hung:2011xb} and furthermore find the correct functional that leads to the constraint equation. This inconsistency between generalized gravity entropy and Jacobson-Myers functional certainly deserves further investigation. We wish to come back to this issue in the future.

Even though generically the Jacobson-Myers functional is not exactly the one for generalized gravitational entropy, it could be a good approximation. As shown in Section 4, the difference between the constraints equations consists of the cubic and higher order terms of the extrinsic curvatures. In the case that such terms could be negligible, one may use the Jacobson-Myers functional to compute the generalized gravitational entropy. However one must justify this approximation case by case.

%If the results in this note is correct, then there would be conclusions as listed above.
In this note, our discussion is based on the approximations in the geometries (\ref{e8}) and (\ref{cone}). However, in the gravity theory with higher derivative terms, such approximation could be too restrictive. It would be very interesting to investigate if the relaxation of the approximation may resolve the puzzle.\footnote{We would like to thank the anonymous referee to inspire this discussion.}

\vspace*{10mm}
\noindent {\large{\bf Acknowledgments}}\\
We would like to thank Juan Maldacena for valuable correspondence, and we also thank Dmitri Fursaev and Aninda Sinha for valuable discussions. BC would like to thank the organizer and participants of the advanced workshop ``Dark Energy and Fundamental Theory" supported by the Special Fund for Theoretical Physics from the National Natural Science Foundations of China with Grant No. 10947203 for stimulating discussions. The work was in part supported by NSFC Grant No. 11275010. JJZ was also in part supported by the Scholarship Award for Excellent Doctoral Student granted by the Ministry of Education of China.
\vspace*{5mm}

\begin{appendix}

\section{Curvatures of the conical geometry}\label{sa}

In this appendix we give the details of the calculation of the curvature tensors of conical geometry (\ref{cone}). The nonvanishing components of the Christoffel connection, Riemann tensor, Ricci tensor and Ricci scalar are
\bea
&& \G^\a_{\b\g}=\p_\b\r\d^\a_\g+\p_\g\r\d^\a_\b-\p^\a\r g_{\b\g},  \nn\\
&& \G^\a_{ij}=-(1-2\r)K^{(\a)}_{ij},  \nn\\
&& \G^i_{j\a}=K_{(\a)j}^{\phantom{(\a)}i},  \nn\\
&& \G^i_{jk}= \g^i_{jk},
\eea
\bea
&& R_{\a\b\g\d}=\p_\b\p_\g\r g_{\a\d}+\p_\a\p_\d\r g_{\b\g}-\p_\b\p_\d\r g_{\a\g}-\p_\a\p_\g\r g_{\b\d},  \nn\\
&& R_{ij\a\b}=K_{(\a)ik}K_{(\b)j}^{\phantom{(\b)}k}-K_{(\a)jk}K_{(\b)i}^{\phantom{(\b)}k},  \nn\\
&& R_{i\a j\b}=K_{(\g)ij}(\p_\a\r\d^\g_\b+\p_\b\r\d^\g_\a-\p^\g\r g_{\a\b})-K_{(\a)jk} K_{(\b)i}^{\phantom{(\b)}k},  \nn\\
&& R_{\a ijk}=D_k K_{(\a)ji}- D_j K_{(\a)ki},  \nn\\
&& R_{ijkl}=\mc R_{ijkl}+(1-2\r)(K_{(\a)il}K^{(\a)}_{jk}-K_{(\a)ik}K^{(\a)}_{jl}),
\eea
\bea
&& R_{\a\b}=-\p^\g\p_\g \r g_{\a\b}+K_{(\g)} (\p_\a\r\d^\g_\b+\p_\b\r\d^\g_\a-\p^\g\r g_{\a\b})-K_{(\a)ij}K_{(\b)}^{ij},  \nn\\
&& R_{\a i}=D_j K_{(\a)i}^{\phantom{(\a)}j}-\p_i K_{(\a)}, \nn\\
&& R_{ij}=\mc R_{ij}-(1-2\r)K^{(\a)}K_{(\a)ij},  \nn\\
&& R=\mc R-2\p^\g\p_\g \r-(1-2\r)(K^{(\a)} K_{(\a)}+K_{{(\a)} ij}K^{(\a) ij}).
\eea
There are some comments for the results. What we mean $\G^i_{jk}= \g^i_{jk}$ is that this components of Christoffel connections given by $g_{\m\n}$ and $h_{ij}$ are the same to the leading order. The symbol $D_i$ means covariant derivative of $h_{ij}$. The curvature tensors $\mc R_{ijkl}$, $\mc R_{ij}$ and $\mc R$ are defined by $h_{ij}$, and so they are the intrinsic curvatures of $\S$. We have focused on the terms to the leading order of $\e$, and the leading terms when $r \to 0$. Also the terms proportional to $\p^\g\p_\g\r$ and $\r$ could be omitted under our approximations.

Besides the curvature $R_{\m\n\r\s}$ defined by $g_{\m\n}$ and the curvature $\mc R_{ijkl}$ defined by $h_{ij}$, there is also the projected curvature $\mf R_{ijkl}$ which is the pullblack of $R_{\m\n\r\s}$ on $\mc M$ to $\S$, and its definition could be written formally as
\be \label{pc}
\mf R_{ijkl} \equiv \p_i X^\m \p_j X^\n \p_k X^\r \p_l X^\s R_{\m\n\r\s},
\ee
with $x^\m=X^\m(y)$ characterizing the embedding of $\S$ in $\mc M$.
For the conical geometry (\ref{cone}) and under our approximations we have
\bea \label{j43}
&& \mf R_{ijkl}=R_{ijkl}=\mc R_{ijkl}+K_{(\a)il}K^{(\a)}_{jk}-K_{(\a)ik}K^{(\a)}_{jl},  \nn\\
&& \mf R_{ik} \equiv h^{jl} \mf R_{ijkl}=\mc R_{ik}+K^{(\a)}_{ij} K_{(a)k}^{\phantom{(a)}j} - K^{(\a)} K_{(\a)ik},  \nn\\
&& \mf R \equiv h^{ik} \mf R_{ik} = \mc R+K^{(\a)}_{ij}K_{(\a)}^{ij}-K^{(\a)}K_{(\a)}.
\eea
It is also useful to write $\a,\b,\cdots$ in the $(z, \bar z)$ coordinates
\bea \label{e20}
&& R_{izjz}=2K_{(z)ij}\p_z \r-K_{(z)ik}K_{(z)j}^{\phantom{(z)}k},  \nn\\
&& R_{i \bar zj \bar z}=2K_{(\bar z)ij}\p_{\bar z} \r-K_{(\bar z)ik} K_{(\bar z)j}^{\phantom{(\bar z)}k},  \nn\\
&& R_{izj\bar z}=-K_{(z)ik} K_{(\bar z)j}^{\phantom{(\bar z)}k},  \nn\\
&& R_{zz}=2K_{(z)}\p_z \r-K_{(z)ij}K_{(z)}^{ij},  \nn\\
&& R_{\bar z\bar z}=2K_{(\bar z)}\p_{\bar z} \r-K_{(\bar z)ij}K_{(\bar z)}^{ij},  \nn\\
&& R_{z\bar z}=-K_{(z)ij}K_{(\bar z)}^{ij}=-\f{1}{4}K_{(\a)ij}K^{(\a)ij}.  \nn\\
\eea

\section{Details of the calculation for Lovelock gravity}\label{sb}

In this section, we present the details of the calculation of (\ref{e39}). Using (\ref{e20}) we have
\be
P_{(m)z}^{\phantom{(m)z}\r\s\l} R_{z\r\s\l}=\f{1}{2}R^{\bar z\r}_{\s\l}P_{(m)z\r}^{\phantom{(m)}\s\l}=
R^{\bar zi}_{zj}P_{(m)zi}^{\phantom{(m)}zj}+
R^{\bar zi}_{\bar zj}P_{(m)zi}^{\phantom{(m)}\bar zj}+
\f{1}{2}R^{\bar zi}_{jk}P_{(m)zi}^{\phantom{(m)}jk}+\cdots.
\ee
We can show that
\bea
&& R^{\bar zi}_{zj}P_{(m)zi}^{\phantom{(m)}zj}=
        \f{m(2m)!}{2^m}
        R^{\bar zi}_{zj}
        \d^{zj \m_2 \n_2 \cdots \m_m \n_m}_{zi \r_2 \s_2 \cdots \r_m \s_m}
        R^{\r_2 \s_2}_{\m_2 \n_2} \cdots R^{\r_m \s_m}_{\m_m \n_m}                                 \nn\\
&& \phantom{R^{\bar zi}_{zj}P_{(m)zi}^{\phantom{(m)}zj}}
  =\f{m(2m)!}{2^m}R^{\bar zi}_{zj}
   \lt( \d^{zj i_2 j_2 \cdots i_m j_m}_{zi k_2 l_2 \cdots k_m l_m}
        R^{k_2 l_2}_{i_2 j_2} \cdots R^{k_m l_m}_{i_m j_m} \rt.                                    \nn\\
&& \phantom{R^{\bar zi}_{zj}P_{(m)zi}^{\phantom{(m)}zj}=\f{m(2m)!}{2^m}R^{\bar zi}_{zj}}
             +4(m-1)R^{\bar z l}_{\bar z k}
             \d^{z\bar z jk i_3 j_3 \cdots i_m j_m}_{z \bar z il k_3 l_3 \cdots k_m l_m}
             R^{k_3 l_3}_{i_3 j_3} \cdots R^{k_m l_m}_{i_m j_m}                                    \\
&& \phantom{R^{\bar zi}_{zj}P_{(m)zi}^{\phantom{(m)}zj}=\f{m(2m)!}{2^m}R^{\bar zi}_{zj}}
      \lt.   +4(m-1)(m-2)R^{mn}_{\bar z k}R^{\bar z l}_{pq}
             \d^{z\bar z jkpq i_4 j_4 \cdots i_m j_m}_{z \bar z imnl k_4 l_4 \cdots k_m l_m}
             R^{k_4 l_4}_{i_4 j_4} \cdots R^{k_m l_m}_{i_m j_m}  \rt),  \nn
\eea

\bea
&& R^{\bar zi}_{\bar zj}P_{(m)zi}^{\phantom{(m)}\bar zj}=
        \f{m(2m)!}{2^m}
        R^{\bar zi}_{\bar zj}
        \d^{\bar zj \m_2 \n_2 \cdots \m_m \n_m}_{zi \r_2 \s_2 \cdots \r_m \s_m}
        R^{\r_2 \s_2}_{\m_2 \n_2} \cdots R^{\r_m \s_m}_{\m_m \n_m}  \nn\\
&& \phantom{R^{\bar zi}_{\bar zj}P_{(m)zi}^{\phantom{(m)}\bar zj}}
        =-\f{m(2m)!}{2^m}
          4(m-1) R^{\bar zi}_{\bar zj}R^{\bar z l}_{z k}
          \d^{z\bar z jk i_3 j_3 \cdots i_m j_m}_{z \bar z il k_3 l_3 \cdots k_m l_m}
          R^{k_3 l_3}_{i_3 j_3} \cdots R^{k_m l_m}_{i_m j_m}  ,
\eea

\bea
&& \f{1}{2}R^{\bar zi}_{jk}P_{(m)zi}^{\phantom{(m)}jk}=
   \f{1}{2} \f{m(2m)!}{2^m}
        R^{\bar zi}_{jk}
        \d^{jk \m_2 \n_2 \cdots \m_m \n_m}_{zi \r_2 \s_2 \cdots \r_m \s_m}
        R^{\r_2 \s_2}_{\m_2 \n_2} \cdots R^{\r_m \s_m}_{\m_m \n_m}  \nn\\
&& \phantom{\f{1}{2}R^{\bar zi}_{jk}P_{(m)zi}^{\phantom{(m)}jk}}
        =\f{m(2m)!}{2^m}
          4(m-1)(m-2) R^{\bar zi}_{jk} R^{\bar z m}_{z l} R^{pq}_{\bar z n}
          \d^{z\bar z jk ln i_4 j_4 \cdots i_m j_m}_{z \bar z impq k_4 l_4 \cdots k_m l_m}
          R^{k_4 l_4}_{i_4 j_4} \cdots R^{k_m l_m}_{i_m j_m},
\eea
and then we have
\be
P_{(m)z}^{\phantom{(m)z}\r\s\l} R_{z\r\s\l}
=\f{m(2m)!}{2^m}R^{\bar zi}_{zj}
        \d^{zj i_2 j_2 \cdots i_m j_m}_{zi k_2 l_2 \cdots k_m l_m}
        R^{k_2 l_2}_{i_2 j_2} \cdots R^{k_m l_m}_{i_m j_m}.
\ee
According to our convention, we have
\bea
&&\d^{zj i_2 j_2 \cdots i_m j_m}_{zi k_2 l_2 \cdots k_m l_m}=\f{1}{2m}\d^{j i_2 j_2 \cdots i_m j_m}_{i k_2 l_2 \cdots k_m l_m}\nn\\
&&\phantom{\d^{zj i_2 j_2 \cdots i_m j_m}_{zi k_2 l_2 \cdots k_m l_m}}=
\f{1}{2m(2m-1)} \lt(  \d^j_i \d^{i_2 j_2 \cdots i_m j_m}_{k_2 l_2 \cdots k_m l_m}
                     -(2m-2) \d^j_{[k_2} \d^{i_2 j_2 \cdots i_m j_m}_{|i| l_2 \cdots k_m l_m]}   \rt).
\eea
Then using (\ref{e20}) and $\mf R_{ijkl}=R_{ijkl}$, we have
\be
P_{(m)z}^{\phantom{(m)z}\r\s\l}R_{z\r\s\l}=-4\p_z \r K_{(z)ij} m
\lt( \mf P_{(m-1)\phantom{i}klp}^{\phantom{(m-1)}i} \mf R^{jklp} -\f{1}{2}\mf L_{(m-1)} h^{ij} \rt)
\ee
with the definition
\be \label{e37}
\mf P_{(m-1)i}^{\phantom{(m-1)i}klp} \equiv \f{\p \mf L_{(m-1)}}{\p \mf R^i_{\phantom{i}klp}}.
\ee

\end{appendix}

%\bibliographystyle{utcaps}     %%非常好，期刊，arXiv超链接
%\bibliographystyle{jhep}    %%好，arXiv超链接
%\bibliographystyle{kp}     %%好，arXiv超链接
%\nocite{*}

%\bibliographystyle{utphys}   %%非常好，期刊，arXiv超链接
%\bibliography{zbib}

\begin{thebibliography}{10}

\bibitem{Bekenstein:1973ur}
J.~D. Bekenstein, ``{Black holes and entropy},''
\href{http://dx.doi.org/10.1103/PhysRevD.7.2333}{{\em Phys.Rev.} {\bfseries D7}
  (1973) 2333--2346}.
%%CITATION = PHRVA,D7,2333;%%.

\bibitem{Hawking:1974sw}
S.~W. Hawking, ``{Particle Creation by Black Holes},''
\href{http://dx.doi.org/10.1007/BF02345020}{{\em Commun.Math.Phys.} {\bfseries
  43} (1975) 199--220}.
%%CITATION = CMPHA,43,199;%%.

\bibitem{Gibbons:1976ue}
G.~Gibbons and S.~Hawking, ``{Action Integrals and Partition Functions in
  Quantum Gravity},''
\href{http://dx.doi.org/10.1103/PhysRevD.15.2752}{{\em Phys.Rev.} {\bfseries
  D15} (1977) 2752--2756}.
%%CITATION = PHRVA,D15,2752;%%.

\bibitem{Lewkowycz:2013nqa}
A.~Lewkowycz and J.~Maldacena, ``{Generalized gravitational entropy},''
\href{http://arxiv.org/abs/1304.4926}{{\ttfamily arXiv:1304.4926 [hep-th]}}.
%%CITATION = ARXIV:1304.4926;%%.

\bibitem{Fursaev:2007sg}
D.~V. Fursaev, ``{Entanglement entropy in quantum gravity and the Plateau
  groblem},'' \href{http://dx.doi.org/10.1103/PhysRevD.77.124002}{{\em
  Phys.Rev.} {\bfseries D77} (2008) 124002},
\href{http://arxiv.org/abs/0711.1221}{{\ttfamily arXiv:0711.1221 [hep-th]}}.
%%CITATION = ARXIV:0711.1221;%%.

\bibitem{Maldacena:1997re}
J.~M. Maldacena, ``{The Large N limit of superconformal field theories and
  supergravity},'' {\em Adv.Theor.Math.Phys.} {\bfseries 2} (1998) 231--252,
\href{http://arxiv.org/abs/hep-th/9711200}{{\ttfamily arXiv:hep-th/9711200
  [hep-th]}}.
%%CITATION = HEP-TH/9711200;%%.

\bibitem{Gubser:1998bc}
S.~Gubser, I.~R. Klebanov, and A.~M. Polyakov, ``{Gauge theory correlators from
  noncritical string theory},''
  \href{http://dx.doi.org/10.1016/S0370-2693(98)00377-3}{{\em Phys.Lett.}
  {\bfseries B428} (1998) 105--114},
\href{http://arxiv.org/abs/hep-th/9802109}{{\ttfamily arXiv:hep-th/9802109
  [hep-th]}}.
%%CITATION = HEP-TH/9802109;%%.

\bibitem{Witten:1998qj}
E.~Witten, ``{Anti-de Sitter space and holography},'' {\em
  Adv.Theor.Math.Phys.} {\bfseries 2} (1998) 253--291,
\href{http://arxiv.org/abs/hep-th/9802150}{{\ttfamily arXiv:hep-th/9802150
  [hep-th]}}.
%%CITATION = HEP-TH/9802150;%%.

\bibitem{Ryu:2006bv}
S.~Ryu and T.~Takayanagi, ``{Holographic derivation of entanglement entropy
  from AdS/CFT},'' \href{http://dx.doi.org/10.1103/PhysRevLett.96.181602}{{\em
  Phys.Rev.Lett.} {\bfseries 96} (2006) 181602},
\href{http://arxiv.org/abs/hep-th/0603001}{{\ttfamily arXiv:hep-th/0603001
  [hep-th]}}.
%%CITATION = HEP-TH/0603001;%%.

\bibitem{Ryu:2006ef}
S.~Ryu and T.~Takayanagi, ``{Aspects of Holographic Entanglement Entropy},''
  \href{http://dx.doi.org/10.1088/1126-6708/2006/08/045}{{\em JHEP} {\bfseries
  0608} (2006) 045},
\href{http://arxiv.org/abs/hep-th/0605073}{{\ttfamily arXiv:hep-th/0605073
  [hep-th]}}.
%%CITATION = HEP-TH/0605073;%%.

\bibitem{Fursaev:2006ih}
D.~V. Fursaev, ``{Proof of the holographic formula for entanglement entropy},''
  \href{http://dx.doi.org/10.1088/1126-6708/2006/09/018}{{\em JHEP} {\bfseries
  0609} (2006) 018},
\href{http://arxiv.org/abs/hep-th/0606184}{{\ttfamily arXiv:hep-th/0606184
  [hep-th]}}.
%%CITATION = HEP-TH/0606184;%%.

\bibitem{Headrick:2010zt}
M.~Headrick, ``{Entanglement Renyi entropies in holographic theories},''
  \href{http://dx.doi.org/10.1103/PhysRevD.82.126010}{{\em Phys.Rev.}
  {\bfseries D82} (2010) 126010},
\href{http://arxiv.org/abs/1006.0047}{{\ttfamily arXiv:1006.0047 [hep-th]}}.
%%CITATION = ARXIV:1006.0047;%%.

\bibitem{Casini:2011kv}
H.~Casini, M.~Huerta, and R.~C. Myers, ``{Towards a derivation of holographic
  entanglement entropy},''
  \href{http://dx.doi.org/10.1007/JHEP05(2011)036}{{\em JHEP} {\bfseries 1105}
  (2011) 036},
\href{http://arxiv.org/abs/1102.0440}{{\ttfamily arXiv:1102.0440 [hep-th]}}.
%%CITATION = ARXIV:1102.0440;%%.

\bibitem{Fursaev:2012mp}
D.~Fursaev, ``{Entanglement Renyi Entropies in Conformal Field Theories and
  Holography},'' \href{http://dx.doi.org/10.1007/JHEP05(2012)080}{{\em JHEP}
  {\bfseries 1205} (2012) 080},
\href{http://arxiv.org/abs/1201.1702}{{\ttfamily arXiv:1201.1702 [hep-th]}}.
%%CITATION = ARXIV:1201.1702;%%.

\bibitem{Jacobson:1993xs}
T.~Jacobson and R.~C. Myers, ``{Black hole entropy and higher curvature
  interactions},'' \href{http://dx.doi.org/10.1103/PhysRevLett.70.3684}{{\em
  Phys.Rev.Lett.} {\bfseries 70} (1993) 3684--3687},
\href{http://arxiv.org/abs/hep-th/9305016}{{\ttfamily arXiv:hep-th/9305016
  [hep-th]}}.
%%CITATION = HEP-TH/9305016;%%.

\bibitem{Wald:1993nt}
R.~M. Wald, ``{Black hole entropy is the Noether charge},''
  \href{http://dx.doi.org/10.1103/PhysRevD.48.R3427}{{\em Phys.Rev.} {\bfseries
  D48} (1993) 3427--3431},
\href{http://arxiv.org/abs/gr-qc/9307038}{{\ttfamily arXiv:gr-qc/9307038
  [gr-qc]}}.
%%CITATION = GR-QC/9307038;%%.

\bibitem{Jacobson:1993vj}
T.~Jacobson, G.~Kang, and R.~C. Myers, ``{On black hole entropy},''
  \href{http://dx.doi.org/10.1103/PhysRevD.49.6587}{{\em Phys.Rev.} {\bfseries
  D49} (1994) 6587--6598},
\href{http://arxiv.org/abs/gr-qc/9312023}{{\ttfamily arXiv:gr-qc/9312023
  [gr-qc]}}.
%%CITATION = GR-QC/9312023;%%.

\bibitem{Iyer:1994ys}
V.~Iyer and R.~M. Wald, ``{Some properties of Noether charge and a proposal for
  dynamical black hole entropy},''
  \href{http://dx.doi.org/10.1103/PhysRevD.50.846}{{\em Phys.Rev.} {\bfseries
  D50} (1994) 846--864},
\href{http://arxiv.org/abs/gr-qc/9403028}{{\ttfamily arXiv:gr-qc/9403028
  [gr-qc]}}.
%%CITATION = GR-QC/9403028;%%.

\bibitem{Iwashita:2006zj}
Y.~Iwashita, T.~Kobayashi, T.~Shiromizu, and H.~Yoshino, ``{Holographic
  entanglement entropy of de Sitter braneworld},''
  \href{http://dx.doi.org/10.1103/PhysRevD.74.064027}{{\em Phys.Rev.}
  {\bfseries D74} (2006) 064027},
\href{http://arxiv.org/abs/hep-th/0606027}{{\ttfamily arXiv:hep-th/0606027
  [hep-th]}}.
%%CITATION = HEP-TH/0606027;%%.

\bibitem{deBoer:2011wk}
J.~de~Boer, M.~Kulaxizi, and A.~Parnachev, ``{Holographic Entanglement Entropy
  in Lovelock Gravities},''
  \href{http://dx.doi.org/10.1007/JHEP07(2011)109}{{\em JHEP} {\bfseries 1107}
  (2011) 109},
\href{http://arxiv.org/abs/1101.5781}{{\ttfamily arXiv:1101.5781 [hep-th]}}.
%%CITATION = ARXIV:1101.5781;%%.

\bibitem{Hung:2011xb}
L.-Y. Hung, R.~C. Myers, and M.~Smolkin, ``{On Holographic Entanglement Entropy
  and Higher Curvature Gravity},''
  \href{http://dx.doi.org/10.1007/JHEP04(2011)025}{{\em JHEP} {\bfseries 1104}
  (2011) 025},
\href{http://arxiv.org/abs/1101.5813}{{\ttfamily arXiv:1101.5813 [hep-th]}}.
%%CITATION = ARXIV:1101.5813;%%.

\bibitem{Ogawa:2011fw}
N.~Ogawa and T.~Takayanagi, ``{Higher Derivative Corrections to Holographic
  Entanglement Entropy for AdS Solitons},''
  \href{http://dx.doi.org/10.1007/JHEP10(2011)147}{{\em JHEP} {\bfseries 1110}
  (2011) 147},
\href{http://arxiv.org/abs/1107.4363}{{\ttfamily arXiv:1107.4363 [hep-th]}}.
%%CITATION = ARXIV:1107.4363;%%.

\bibitem{Headrick:2007km}
M.~Headrick and T.~Takayanagi, ``{A Holographic proof of the strong
  subadditivity of entanglement entropy},''
  \href{http://dx.doi.org/10.1103/PhysRevD.76.106013}{{\em Phys.Rev.}
  {\bfseries D76} (2007) 106013},
\href{http://arxiv.org/abs/0704.3719}{{\ttfamily arXiv:0704.3719 [hep-th]}}.
%%CITATION = ARXIV:0704.3719;%%.

\bibitem{Guo:2013aca}
W.-z. Guo, S.~He, and J.~Tao, ``{Note on Entanglement Temperature for Low
  Thermal Excited States in Higher Derivative Gravity},''
\href{http://arxiv.org/abs/1305.2682}{{\ttfamily arXiv:1305.2682 [hep-th]}}.
%%CITATION = ARXIV:1305.2682;%%.

\bibitem{Bhattacharya:2012mi}
J.~Bhattacharya, M.~Nozaki, T.~Takayanagi, and T.~Ugajin, ``{Thermodynamical
  Property of Entanglement Entropy for Excited States},'' {\em Phys. Rev. Lett.
  110,} {\bfseries 091602} (2013) ,
\href{http://arxiv.org/abs/1212.1164}{{\ttfamily arXiv:1212.1164 [hep-th]}}.
%%CITATION = ARXIV:1212.1164;%%.

\bibitem{Nozaki:2013vta}
M.~Nozaki, T.~Numasawa, A.~Prudenziati, and T.~Takayanagi, ``{Dynamics of
  Entanglement Entropy from Einstein Equation},''
\href{http://arxiv.org/abs/1304.7100}{{\ttfamily arXiv:1304.7100 [hep-th]}}.
%%CITATION = ARXIV:1304.7100;%%.

\bibitem{Bhattacharyya:2013jma}
A.~Bhattacharyya, A.~Kaviraj, and A.~Sinha, ``{Entanglement entropy in higher
  derivative holography},''
\href{http://arxiv.org/abs/1305.6694}{{\ttfamily arXiv:1305.6694 [hep-th]}}.
%%CITATION = ARXIV:1305.6694;%%.

\bibitem{Solodukhin:1994yz}
S.~N. Solodukhin, ``{The Conical singularity and quantum corrections to entropy
  of black hole},'' \href{http://dx.doi.org/10.1103/PhysRevD.51.609}{{\em
  Phys.Rev.} {\bfseries D51} (1995) 609--617},
\href{http://arxiv.org/abs/hep-th/9407001}{{\ttfamily arXiv:hep-th/9407001
  [hep-th]}}.
%%CITATION = HEP-TH/9407001;%%.

\bibitem{Fursaev:1995ef}
D.~V. Fursaev and S.~N. Solodukhin, ``{On the description of the Riemannian
  geometry in the presence of conical defects},''
  \href{http://dx.doi.org/10.1103/PhysRevD.52.2133}{{\em Phys.Rev.} {\bfseries
  D52} (1995) 2133--2143},
\href{http://arxiv.org/abs/hep-th/9501127}{{\ttfamily arXiv:hep-th/9501127
  [hep-th]}}.
%%CITATION = HEP-TH/9501127;%%.

\bibitem{Azeyanagi:2007bj}
T.~Azeyanagi, T.~Nishioka, and T.~Takayanagi, ``{Near Extremal Black Hole
  Entropy as Entanglement Entropy via AdS(2)/CFT(1)},''
  \href{http://dx.doi.org/10.1103/PhysRevD.77.064005}{{\em Phys.Rev.}
  {\bfseries D77} (2008) 064005},
\href{http://arxiv.org/abs/0710.2956}{{\ttfamily arXiv:0710.2956 [hep-th]}}.
%%CITATION = ARXIV:0710.2956;%%.

\bibitem{Myers:2010tj}
R.~C. Myers and A.~Sinha, ``{Holographic c-theorems in arbitrary dimensions},''
  \href{http://dx.doi.org/10.1007/JHEP01(2011)125}{{\em JHEP} {\bfseries 1101}
  (2011) 125},
\href{http://arxiv.org/abs/1011.5819}{{\ttfamily arXiv:1011.5819 [hep-th]}}.
%%CITATION = ARXIV:1011.5819;%%.

\bibitem{Unruh:1989hy}
W.~G. Unruh, G.~Hayward, W.~Israel, and D.~Mcmanus, ``{COSMIC STRING LOOPS ARE
  STRAIGHT},''
\href{http://dx.doi.org/10.1103/PhysRevLett.62.2897}{{\em Phys.Rev.Lett.}
  {\bfseries 62} (1989) 2897--2900}.
%%CITATION = PRLTA,62,2897;%%.

\bibitem{Boisseau:1996bp}
B.~Boisseau, C.~Charmousis, and B.~Linet, ``{Dynamics of a selfgravitating thin
  cosmic string},'' \href{http://dx.doi.org/10.1103/PhysRevD.55.616}{{\em
  Phys.Rev.} {\bfseries D55} (1997) 616--622},
\href{http://arxiv.org/abs/gr-qc/9607029}{{\ttfamily arXiv:gr-qc/9607029
  [gr-qc]}}.
%%CITATION = GR-QC/9607029;%%.

\bibitem{Hubeny:2007xt}
V.~E. Hubeny, M.~Rangamani, and T.~Takayanagi, ``{A Covariant holographic
  entanglement entropy proposal},''
  \href{http://dx.doi.org/10.1088/1126-6708/2007/07/062}{{\em JHEP} {\bfseries
  0707} (2007) 062},
\href{http://arxiv.org/abs/0705.0016}{{\ttfamily arXiv:0705.0016 [hep-th]}}.
%%CITATION = ARXIV:0705.0016;%%.

\bibitem{Bunch:1981}
T.~S. Bunch, ``{Surface terms in higher derivative gravity},''
  \href{http://dx.doi.org/10.1088/0305-4470/14/5/008}{{\em J.Phys.} {\bfseries
  A14} (1981) L139--L143}.

\bibitem{Myers:1987yn}
R.~C. Myers, ``{Higher-derivative gravity, surface terms, and string theory},''
\href{http://dx.doi.org/10.1103/PhysRevD.36.392}{{\em Phys.Rev.} {\bfseries
  D36} (1987) 392}.
%%CITATION = PHRVA,D36,392;%%.

\bibitem{Solodukhin:2008dh}
S.~N. Solodukhin, ``{Entanglement entropy, conformal invariance and extrinsic
  geometry},'' \href{http://dx.doi.org/10.1016/j.physletb.2008.05.071}{{\em
  Phys.Lett.} {\bfseries B665} (2008) 305--309},
\href{http://arxiv.org/abs/0802.3117}{{\ttfamily arXiv:0802.3117 [hep-th]}}.
%%CITATION = ARXIV:0802.3117;%%.

\end{thebibliography}

\providecommand{\href}[2]{#2}\begingroup\raggedright\endgroup

\end{document}